\documentclass[aps,pre,twocolumn,superscriptaddress,reprint]{revtex4-2}
\usepackage{dcolumn}
\usepackage{array,epsfig,graphicx,color}
\usepackage{newtxtext}
\usepackage{comment}
\usepackage{amsmath,amssymb,amsfonts,amscd}
\renewcommand{\d}{\mathrm d}

\newcommand{\nn}{\nonumber}
\newcommand{\be}{\begin{equation}}
\newcommand{\ee}{\end{equation}}

\newcommand{\lpa}{\left\langle}
\newcommand{\rpa}{\right\rangle}
\newcommand{\kB}{k_{_\text{B}}}
\usepackage[dvipsnames]{xcolor}

\begin{document}
\title{Memory-aware feedback enhances power in active information engines}
\author{Sehoon Bahng}%
\affiliation{Department of Physics, Ulsan National Institute of Science \& Technology, Ulsan 44919, Korea}
\author{Jae Sung Lee}%
\email{jslee@kias.re.kr}
\affiliation{School of Physics and Quantum Universe Center, 
Korea Institute for Advanced Study, Seoul 02455, Korea}
\author{Cheol-Min Ghim}%
\email{cmghim@unist.ac.kr}%
\affiliation{Department of Physics, Ulsan National Institute of Science \& Technology, Ulsan 44919, Korea}
\affiliation{Department of Biomedical Engineering, 
Ulsan National Institute of Science \& Technology, Ulsan 44919, Korea}
\affiliation{UNIST Research Center for Soft and Living Matter, Ulsan 44919, Korea}

\date{\today}

\begin{abstract}
We study an information engine operating in an active bath, where a Brownian particle confined in a harmonic trap undergoes feedback-driven displacement cycles. Unlike thermal environments, active baths exhibit temporally correlated fluctuations, introducing memory effects that challenge conventional feedback strategies. Extending the framework of stochastic thermodynamics to account for such memory, we analyze a feedback protocol that periodically shifts the potential minimum based on noisy measurements of the particle’s position.
We show that conventional feedback schemes, optimized for memoryless thermal baths, can degrade performance in active media due to the disruption of bath-particle memory by abrupt resetting. To overcome this degradation, we introduce a class of memory-preserving feedback protocols that partially retain the covariance between the particle’s displacement and active noise, thereby exploiting the temporal persistence of active fluctuations.
Through asymptotic analysis, we show how the feedback gain—which quantifies the strength of positional shifts—nontrivially shapes the engine’s work and power profiles. In particular, we demonstrate that in  active media, intermediate gains outperform full-shift resetting.
Our results reveal the critical interplay between bath memory, measurement noise, and feedback gain, offering guiding principles for designing high-performance information engines in nonequilibrium environments.
\end{abstract}

\maketitle

\section{Introduction}
Since Maxwell’s thought experiment\,\cite{maxwell.1871,knott.1911} and Szilard’s design of an engine that converts information into work\,\cite{szilard.1929}, the relationship between information and thermodynamic work has driven foundational advances in nonequilibrium statistical physics. What began as a conceptual paradox has evolved into the rigorous framework of \textit{information engines}, which combine principles from statistical mechanics, information theory, and thermodynamics of small systems.

A cornerstone of this development is Landauer’s principle, which states that erasing one bit of information incurs an irreducible energetic cost\,\cite{landauer.1961,maruyama.2009}. This insight resolved the apparent violation of the second law and established a framework for fully accounting for the thermodynamics of information processing. Within this context, information engines emerged as key model systems where measurement and feedback enable the extraction of work from fluctuations. Experimental realizations, ranging from Szilard-type setups to feedback-controlled optical traps, have confirmed that information can serve as a thermodynamic resource\,\cite{sagawa.2010,sagawabook.2012,blickle.2012,sagawa.2012,parrondo.2015,lutz.2015,martinez.2016,koski.2014,goerlich.2025}.

More recently, attention has shifted to engines operating in nonequilibrium environments, where microscopic energy injection gives rise to persistent, non-thermal fluctuations. Active matter systems, comprising self-propelled colloids, bacteria, and synthetic microswimmers, serve as a prime example\,\cite{zottl.2016,paneru.2018,obyrne.2022}. These systems exhibit intrinsic memory and violate detailed balance, presenting new challenges and opportunities for information thermodynamics. In such active media, can work still be reliably extracted? Do the nonequilibrium statistics of active fluctuations enhance or hinder performance\,\cite{Krish2016NatPhys,lee2020brownian}? And can engines exploit temporal correlations to surpass classical bounds\,\cite{holubec.2020,albay.2023,paneru.2022}?

Recent theoretical studies suggest that active baths can indeed serve as effective energy reservoirs\,\cite{maggi.2014,paneru.2022,paneru.2023}. However, standard feedback protocols—typically designed for memoryless thermal systems—often fail to capitalize on the temporal structure of active fluctuations. While intrinsic memory is commonly viewed as a complicating factor, it also offers a potentially exploitable resource.

In this work, we develop a stochastic thermodynamic framework for an information engine operating in an active bath with temporally correlated noise. 
We propose a proportional adjustment feedback protocol that partially retains the system’s memory, allowing feedback to better align with the underlying dynamics. 
Using a harmonic trap under overdamped Langevin dynamics, we show that this memory-aware control strategy enhances work extraction compared to conventional feedback protocol.
Our results reveal that temporal correlations can be harnessed to improve engine performance.
This study extends the scope of information thermodynamics of active and temporally structured environments by highlighting the utility of adaptive, memory-aware feedback in nonequilibrium control.

The paper is organized as follows. In Section~\ref{Sec:Model}, we describe the setup of the information engine, consisting of a Brownian particle confined in a harmonic trap and immersed in an active bath. The engine is operated using measurements with finite error and proportional-shift feedback. In Section~\ref{Sec:Control}, we investigate how the proportional-shift feedback preserves the memory of the active system during relaxation dynamics of the engine. Section~\ref{Sec:Optimal} highlights the advantages of this memory-aware feedback in two aspects. First, in Section~\ref{Sec: SNR}, we show that the proportional-shift feedback scheme extracts more work under erroneous measurements. Second, in Section~\ref{Sec: maxpower}, we demonstrate that it enables higher power extraction compared to the conventional feedback scheme. Finally, we conclude in Section~\ref{Sec:Conclusion}.

\section{Model} \label{Sec:Model}
We consider a detector, modeled as a Maxwell's demon, that measures the position of a Brownian particle immersed in an active bath. In addition to experiencing viscous damping, the particle is subject to both thermal noise 
$\xi$ and active noise $\eta$. If the demon also confines the particle within a harmonic potential, the particle's dynamics in the overdamped regime is governed by the following equation of motion:
\be
\gamma\dot{x}_t = -\kappa(x_t-\lambda_t) + \xi_t + \eta_t,
\label{eq_LanEq}
\ee
where the overdot denotes time derivative. The parameters $\gamma$, $\kappa$ and $\lambda$ represent the damping constant, the stiffness of the harmonic trap, and the center of potential, respectively.  All variables with subscript $t$ evolve in time. Denoting ensemble averages by angular brackets $\langle\cdot\rangle$, thermal noise satisfies $\langle \xi_t \rangle = 0$, $\langle \xi_t \xi_{t'} \rangle = 2\gamma k_{_\text{B}}T \delta(t-t')$ where $k_{_\text{B}}$ is the Boltzmann constant and $T$ is the temperature of the thermal bath, consistent with the Einstein relation. The active noise satisfies $\langle \eta_t \rangle = 0$, $\langle\eta_t\eta_{t'}\rangle=A^2\exp(-|t'-t|/\tau_c)$, where $A$ and $\tau_c$ are the parameters representing the amplitude and correlation time of the noise, respectively\,\cite{argun.2016,maggi.2014}.  The thermal and active noises are statistically independent, as ensured by $\langle \xi_t \eta_{t^\prime} \rangle =0$.
\begin{figure}[t!] 
\centerline{
{\resizebox{\columnwidth}{!}{\includegraphics{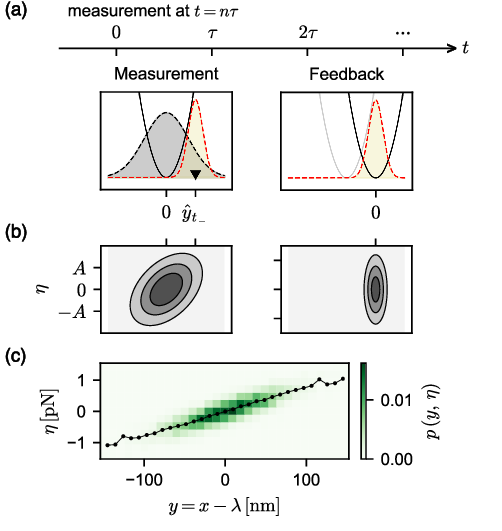}}}}
\caption{(Color online) Operation of an information engine and memory effect in active baths. 
(a) Along the time line, measurement and feedback occur at $t = n\tau_-$ and $t = n\tau_+$, respectively. 
In the left panel (Measurement), the inverted triangle indicates the measured position $\hat y_{t_-}$ relative to the center of the potential.
The red dashed line represents the particle density inferred from the measurement. 
In the right panel (Feedback), an instantaneous shift of the potential center changes the reference frame. 
(b) The left and right panels display contour plots of the joint probability distribution between the relative position $y$ and the active noise $\eta$ at $n \tau_-$ and $n \tau_+$, respectively. 
The distributions are evaluated in the periodic steady state. 
(c) The heatmap shows a joint probability distribution $p(y_t,\eta_t)$ obtained from numerical simulations using active Ornstein-Uhlenbeck (AOUP) process. 
The black dots and the connecting line represents the conditional average of the active noise given the position $y$.}
\label{fig-model}
\end{figure}

Unlike the bath-induced trajectories $\xi_t$ and $\eta_t$, the trajectory of $\lambda_t$ is controlled by the demon based on its observation $\hat{x}_t$. The measurement error, defined as $\epsilon_t = \hat{x}_t - x_t$, is not accessible to the demon. In some experimental setups, the demon only observes the relative position of the particle with respect to the potential center, $y_t=x_t-\lambda_t$. In such cases, the corresponding measurement error, $\hat{y}_t-y_t$ coincides with the error in the absolute frame, $\epsilon_t$.

We focus on a preliminary ansatz for an information engine, where measurements are made discretely at finite time intervals $\tau$. Immediately after each measurement, the demon shifts the potential to the desired location within a short duration $\tau_\text{FB}$. This target location depends on the experimental goal, such as unidirectional ratcheting\,\cite{paneru.2018, saha.2021} or power extraction\,\cite{paneru.2020, paneru.2022}.

To simplify the feedback protocol, we consider the idealized limit, in which the potential shift occurs instantaneously, i.e., $\tau_\textrm{FB} =0$. In this limit, the entire change in potential energy is converted into work, as there is no time for heat dissipation. This simplification has been used in previous studies\,\cite{abreu.2011}. With all of the above preparations, the extracted work is given by
\be
W_t = \frac{1}{2} \kappa\left(y_{t_-}^2-y_{t_+}^2\right),
\label{eq_StocWork}
\ee 
where $t_\pm$ denotes the time point immediately before ($-$) and after ($+$) the potential is shifted, as illustrated in Fig.~\ref{fig-model}(a). Note that $y_{t_-}$ and $y_{t_+}$ exhibit a discontinuous jump, in contrast to $x_t$, which remains uniquely defined. Positive (negative) $W_t$ denotes work extraction from (injection into) the particle.

If the measurement is free from error, the protocol $\lambda_{t_+}=\hat{x}_t$~(i.e.,~$\hat{y}_t=0$) clearly yields the most exploitative update, and this has been used as a practical experimental ansatz\,\cite{paneru.2018}. On the other hand, when the measurement error is present, the Bayesian estimate of $y_t$ offers a more refined update compared to the direct use of raw measurement\,\cite{saha.2022}. 
In active baths, feedback based on error-free measurements has been assumed to be the most exploitative protocol\,\cite{paneru.2022, saha.2023}. However, this assumption breaks down as such feedback disrupts memory in active baths.

Quantifying memory in a meaningful way requires consideration of both the system and its intended function\,\cite{zhao2024emergence,paga2024quantifying}. 
In our case, we focus on the expectation value of $W_t$, determined by $\langle y_t^2\rangle$, while analysis of higher moments remains a natural and important direction, particularly in connection with fluctuation theorems \cite{paneru.2020, paneru.2022}.
During the relaxation process, the statistics of $x_t$ is coupled with those of the active noise $\eta_t$. However, this correlation is erased immediately after the feedback, as illustrated in Fig.~\ref{fig-model}(b).

More specifically, the relaxation dynamics are influenced by the covariance between the Brownian particle’s position and the active noise. Feedback interrupts this relaxation by resetting the covariance to zero. Position-correlated active noise enhances diffusion, but this enhancement is destroyed by feedback. Although Fig.~\ref{fig-model}(b) specifically considers Gaussian-colored noise, such as that of the active Ornstein-Uhlenbeck process (AOUP)\,\cite{martin.2021}, we will later show that the covariance dynamics are consistent across various types of active noise. 

Fig.~\ref{fig-model}(c) shows the joint probability distribution $p(y_t,\eta_t)$ as it relaxes toward a nonequilibrium steady state.
For AOUP, the mean active noise for a given position $y$ is given as follows\,\cite{holt2023essential}:
\be
\bar{\eta}_t(y) \equiv \int \eta_t P(\eta_t | y) \d \eta_t = \frac{\langle y_t \eta_t \rangle}{\langle y_t^2 \rangle}y.
\label{eq-effective_field}
\ee
However, such compact calculation of Eq.~(\ref{eq-effective_field}) is not always achievable if the active noise is non-Gaussian~\cite{dhar2019run}. 

Finally, we propose a proportional-shift feedback scheme to mitigate memory loss during the feedback process. As illustrated in Fig.~2(a), the feedback is implemented as
\be
\lambda_{t_+}-\lambda_{t_-}=g\left(\hat{x}_t - \lambda_{t_-}\right)= g \hat{y}_{t_-},
\label{eq_PropFeed}
\ee
where $g$ is the proportional gain parameter, with the standard
full-shift feedback (FSF) being recovered when $g=1$.
Similar protocols have previously been proposed in different contexts,
for example, to generate a broad family of nonequilibrium steady states via partial 
resetting\,\cite{olsen2024thermodynamic}, or to test thermodynamic uncertainty relations 
in thermal baths\,\cite{paneru2020reaching}. In the present work, we investigate how modulating the feedback influences its interplay with the system's relaxation dynamics, and how this, in turn, impacts work extraction.

\section{Control of relaxation dynamics: Memory-assisted relaxation} \label{Sec:Control}
This section investigates the relaxation dynamics governed by proportional adjustment feedback as the system evolves from a transient to a steady state under cyclic operation. This analysis leads to an understanding of how memory-preserving feedback enhances the work extraction capability of the information engine.

\begin{figure}[t] 
\centerline{
{\resizebox{\columnwidth}{!}{\includegraphics{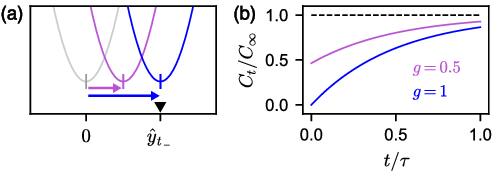}}}}
\caption{(Color online) Proportional adjustment as a protocol for memory preservation. 
(a) The inverted triangle indicates the measured position of the Brownian particle 
at time $t$, and the horizontal arrows denote the shift of the potential center, 
with magenta and blue vertical markers representing proportional adjustment and 
full shift, respectively. (b) Time evolution of the covariance $C_t$ between 
the relative position $y_t$ and the active noise $\eta_t$: blue for $g=1$ and 
magenta for $g=0.5$. The initial  covariance is taken from the steady state of the cyclic operation, and increases as the proportionality $g$ deviates from 1. 
The black dashed line denotes the asymptotic covariance in the limit $\tau\to\infty$.}
\label{fig-relaxation}
\end{figure}

Consider a stochastic trajectory with an arbitrary initial condition $y_{0_+}$. From Eq. (\ref{eq_LanEq}), the state at the end of the relaxation period is given by
\be
y_{\tau_-}=y_{0_+}e^{-\tau/\tau_r} + \frac{1}{\gamma} e^{-\tau/\tau_r} \int_{0_+}^{\tau_-} \d t \, e^{t/\tau_r} (\xi_t + \eta_t),
\label{eq-TrajRelx}
\ee
where $\tau_r\!\equiv\!\gamma/\kappa$ denotes the characteristic relaxation timescale. The demon measures the position of the particle, but the measurement error $\epsilon_\tau$ remains unknown to it. Following Eq. (\ref{eq_PropFeed}), the demon updates the center of the potential, resulting in 
\be
y_{\tau_+}= (1-g)y_{\tau_-} -g\epsilon_{\tau}~.
\label{eq-TrajFeed}
\ee
By substituting $y_{\tau_-}$ in Eq.~(\ref{eq-TrajFeed}) with Eq.~(\ref{eq-TrajRelx}), one can express $y_{\tau_+}$, and consequently the entire sequence $\{y_{t_+}|\,t\!=\!n\tau, n\!\in\!\mathbb{N}\}$, in terms of $y_{0_+}$. Specifically, the relative mean position of the particle evolves as 
\be
\langle y_{n\tau_+}\rangle=\langle y_{0_+}\rangle(1-g)^n e^{-n\tau/\tau_r},
\label{yinfty}
\ee
which converges as $n\to\infty$ only if $\lvert (1-g)e^{-\tau/\tau_r}\rvert < 1$. This condition leads to the following requirement on $\tau$, that is,
\be
\tau>
\begin{cases}
~0~, & \text{if}~0 < g < 2 \\
~\tau_r \ln \lvert 1-g \rvert~, & \text{otherwise.}
\end{cases}
\label{eq-CondSteady}
\ee

The time evolution of the variance $\langle y_t^2\rangle-\langle y_t\rangle^2\equiv V_t$ is less straightforward than that of the mean, due to the presence of the covariance $C_t\equiv\langle y_{t}\eta_t \rangle$ between the relative position of the particle and the active noise. From Eq. (\ref{eq-TrajRelx}) and Appendix~\ref{App_covariance_dynamics},
\begin{align}
V_{\tau_-} &=V_{0_+}e^{-2\tau/\tau_r} + \kappa^{-1}\left(k_{_\text{B}}T+C_\infty\right)\left(1-e^{-2\tau/\tau_r}\right) \nn \\ 
&~-2\kappa^{-1}(C_\infty-C_{0_+})\frac{e^{-2\tau/\bar{\tau}}-e^{-2\tau/\tau_r} }{1-\tau_r/\tau_c}, 
\label{eq-VarRelx}
\end{align}
where $\bar{\tau} \equiv 2\tau_c \tau_r /(\tau_c + \tau_r)$, and the constant $C_\infty=\lim_{\tau\to\infty} C_{\tau_-}=\bar{\tau} A^2/2\gamma$ characterizes the steady-state covariance. The two terms $V_{0_+}$ and $C_{0_+}$ are determined from the initial distribution of $y_{0_+}$. Eq. (\ref{eq-VarRelx}) also holds in the case $\tau_r=\tau_c$ when interpreted in the limiting sense. In general, the time evolution of the covariance is given by
\be
C_{t} = C_{0_+} e^{-2 t/\bar{\tau}} 
+C_\infty\left(1-e^{-2t/\bar{\tau}}\right)~.
\label{eq-CovRelx}
\ee
Note that Eqs.~(\ref{eq-VarRelx}) and (\ref{eq-CovRelx}) are generally valid for various types of active noise. In the absence of active noise ($A=0$), $C_t$ is simply zero, and the right-hand side of Eq.~(\ref{eq-VarRelx}) reduces to the case of a Brownian particle in the thermal bath, leaving only the thermal contribution $k_{_\text{B}}T/\kappa\equiv V_\xi$.

Under the feedback protocol with measurement error, Eq.~\eqref{eq-TrajFeed} leads to the following update rules for the (co)variance: 
\begin{subequations}
\begin{align}
V_{\tau_+}&=(1-g)^2 V_{\tau_-} + g^2\sigma^2 \label{eq-VarFeed} \\
C_{\tau_+}&=(1-g) C_{\tau_-}, \label{eq-CovFeed}
\end{align}
\end{subequations}
where $\sigma$ is the standard deviation of the measurement error $\epsilon$.
The sequences $V_{n\tau_+}$ and $C_{n\tau_+}$  $(n=1,2,\ldots)$ converge to their steady-state values as $n\to\infty$, leading to the conditions $V_{\tau_+}=V_{0_+}$ and $C_{\tau_+}=C_{0_+}$. Although their closed-form analytical expressions are difficult to obtain, the time evolution of $C_t$ is illustrated in Fig. \ref{fig-relaxation}(b) for representative feedback gains $g=1$ and $g=0.5$. FSF completely resets the covariance to zero after each cycle. In contrast, attenuated feedback with $0<g<1$ leads to a strictly positive covariance that increases monotonically over time.

In summary, by modulating the feedback adjustment gain,
the engine’s cyclic steady state can be replaced with a controllable,
switchable relaxation behavior. Moreover, the cyclic dynamics 
determine the average work extracted per cycle:
\be 
\langle W_\text{cyc} \rangle = \dfrac{\kappa}{2} \left( V_{\tau_-} - V_{\tau_+} \right). 
\label{avg_work} 
\ee
Importantly, Eq. \eqref{avg_work} depends solely on covariance and do not require higher-order moment contributions.
Thus, covariance serves as a complete and sufficient metric to quantify the relationship between memory retention and the power output of an information engine.

\section{Optimal performance metrics in active baths} \label{Sec:Optimal}

This section examines how the engine’s average work output depends on the properties of the surrounding bath and the feedback gain. It further highlights the engine’s robustness under significant measurement errors, showing that effective work extraction can still be achieved through proportional adjustments to the feedback gain.

\begin{figure}[t!] 
\centerline{
{\resizebox{\columnwidth}{!}{\includegraphics{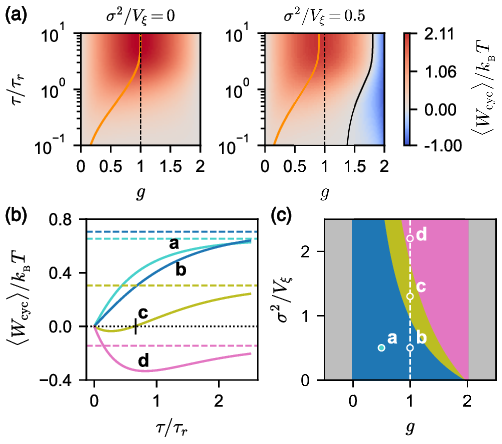}}}}
\caption{(Color online) Profiles of average work extraction per cycle and its classification under varying control parameters. (a) Heatmap of the average work extracted per cycle as a function of the cycle period $\tau$ and feedback gain $g$. The black solid line in the right panel marks the zero-work boundary. The orange curve on each panel indicates the optimal feedback gain $g*$ for each value of $\tau$. As the cycle period shortens, $g*$ decreases monotonically below 1, deviating from the standard full-shift feedback (FSF). (b) Average extracted work per cycle as a function of $\tau$ for different combinations of feedback gain and measurement error: cyan, blue, olive-green, and magenta curves correspond to $(g,\sigma^2/V_\xi) = (0.5,~0.5)$, $(1.0,~0.5)$, $(1.0,~1.3)$, and $(1.0,~2.2)$, respectively, all in identical active baths. For the olive-green curve, the black vertical marker indicates the transition point where the average work changes sign, near $\tau \sim 0.66\tau_r$. Dashed lines indicate the asymptotic values as $\tau \to \infty$.
(c) Classification of operational regimes in the $(g,\sigma^2)$ parameter space. The blue, magenta, and olive-green regions indicate parameter combinations where the average work is always positive, always negative, or changes sign as a function of $\tau$, respectively. The gray shaded region corresponds to settings where the engine either fails to satisfy the convergence condition of Eq.~(\ref{eq-CondSteady}) or yields consistently negative work. }
\label{fig-work}
\end{figure}

\subsection{Work extraction efficiency under noisy feedback} \label{Sec: SNR}
For error-free measurements, the sign of extracted work is straightforward. When the feedback gain is $g=0$ or $2$, no work is extracted in any cycle. For feedback with $0<g<2$, each cycle extracts non-negative work. Conversely, for $g<0$ or $g>2$, no cycle extracts positive work. 

However, when measurement error is present, both positive and negative work extractions may occur within individual cycles. Combining Eq. \eqref{eq-VarFeed} with Eq. \eqref{avg_work}, we obtain
\be 
\langle W_{\text{cyc}} \rangle =\frac{\kappa}{2}g\left[2V_{\tau_-}-g(V_{\tau_-}+\sigma^2)\right]~. 
\label{eq-AvgWork} 
\ee
From Eq.~\eqref{eq-AvgWork}, it is straightforward to see that $\langle W_{\rm cyc}\rangle$ is always negative when $g>2$ or $g<0$. Therefore, in the following discussion, we restrict our attention to the range $0<g<2$, where the system can potentially operate as an engine through the interplay of the control parameters: the feedback gain $g$, the cycle period $\tau$, and the measurement error $\sigma$. For a given value of $V_{\tau_-}$, the extracted work is maximized when the feedback gain is set to 
\be
g_\text{opt}=\frac{1}{1+\sigma^2/V_{\tau_-}}, 
\label{eq-opt_cond}
\ee
which is a self-consistent expression, since $V_{\tau_-}$ itself retains a dependence on $g$ through its evolution from the previous feedback cycle. 

In the long-period limit ($\tau\gg\tau_r$), this optimal gain reduces to $g_\text{opt}=1$ in the absence of measurement error ($\sigma=0$). However, when a measurement error is present, the optimal gain becomes attenuated, that is, $g_\text{opt}<1$, as shown in Fig. \ref{fig-work}(a).
In this limit, the average extracted work per cycle approaches
\be
\lim_{\tau\to\infty}\langle W_\text{cyc} \rangle
=g\left(1-\frac{g}{2}\right)\kappa V_\infty-\frac{g^2}{2}\kappa\sigma^2,
\label{eq-work_slow_limit}
\ee
where $V_\infty\equiv(k_{_\text{B}}T+C_\infty)/\kappa$. In the short-$\tau$ limit ($\tau\ll\tau_r$), from the analysis in the Appendix B, $\lim_{\tau\to 0}\langle W_\text{cyc} \rangle=0$ and
\be
\frac{\d}{\d\tau}\langle W_\text{cyc} \rangle \bigg|_{\tau\to 0}
=\frac{\kappa}{\tau_r}\left(V_\xi-\sigma^2\right)~.
\label{zeropower}
\ee 
This expression reveals that, in the short-period limit, the sign of the extracted work is governed solely by the interplay between thermal fluctuations and measurement error, and remains independent of active noise.

The behavior of work per cycle as a function of the period $\tau$ is illustrated in Fig.~\ref{fig-work}(b). Notably, the transition in the sign of extracted work with varying period is a unique feature of active baths, which is absent in purely thermal environments. This sign-changing regime arises when the measurement error satisfies 
\be
(2/g-1)V_\xi < \sigma^2 \leq (2/g-1)V_\infty,
\label{eq-work_nonmonotonic}
\ee
as indicated by the olive-green region in Fig. \ref{fig-work}(c). 

For amplified feedback ($g>1$), this transition arises at lower levels of measurement noise, as shown in Fig.~\ref{fig-work}(c). In contrast, attenuated feedback ($0<g<1$) allows positive work extraction even under high measurement error, underscoring its robustness to noise. 

In the absence of explicit knowledge of the measurement error, the analysis can be generalized by considering its possible range. The maximum conceivable error then provides a conservative bound on the feedback gain that should not be exceeded to ensure positive work extraction. For instance, if positivity is required for all values of $\tau$, the corresponding maximal feedback gain is determined by the boundary between the olive-green and blue regions in Fig.~3(c).

\begin{figure}[t!] 
\centerline{
{\resizebox{\columnwidth}{!}{\includegraphics{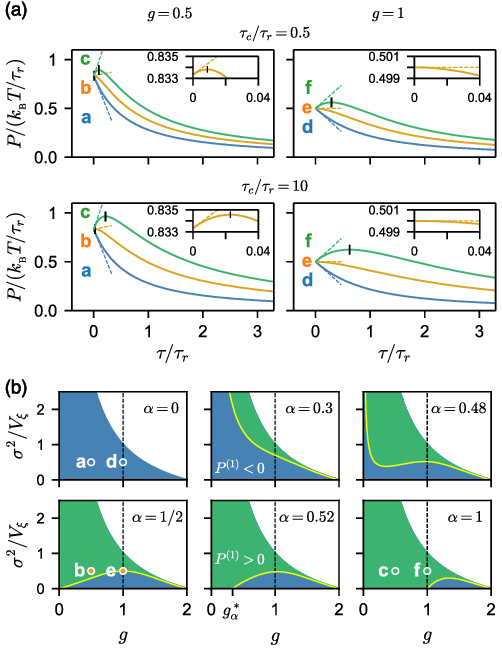}}}}
\caption{(Color online) Profiles of power and classification under varying control parameters. (a) Power profiles are shown for different activity levels, defined by the dimensionless parameter $\alpha\equiv A^2/2\kappa k_{\text{B}}T$, represented by blue, orange, and green curves. Dashed lines indicate linear fits in the short-period limit ($\tau\ll\tau_r$), defined by the intercept $P^{(0)}$ and slope $P^{(1)}$. Each row shares a common active noise correlation time. The left column displays results for FSF, while the right column shows those for an attenuated feedback ($g=0.5$). Insets zoom into the short-period regime of the orange curves, and vertical black lines indicate the period $\tau$ at which the power is maximized.
(b) All six panels include the same boundary curves, which demarcate regions where the short-period intercept $P^{(0)}$ is positive (lower left) or negative (upper right). Yellow curves indicate the activity-dependent boundary where the short-period slope $P^{(1)}$ changes sign, positive in the green/white regions and negative in the blue regions.
}
\label{fig-maxpower_phase2}
\end{figure}

\subsection{Power profiles under noisy feedback} \label{Sec: maxpower}
The performance of an information engine is often evaluated by its power, defined as the average extracted work per unit time, $P=\langle W_\text{cyc}\rangle/\tau$.
Fig.~\ref{fig-maxpower_phase2}(a) shows the power profile as a function of the cycle period $\tau$. As $\tau\to 0$, power approaches a finite value; as $\tau \to \infty$, it decays to zero due to the boundedness of $\langle W_\text{cyc} \rangle$.

In the slow-cycle regime ($\tau\gg\tau_r$) power asymptotically behaves as (see Appendix B)
\be
P \simeq \mathcal{P}^{(0)} + \mathcal{P}^{(1)}\frac{\tau_r}{\tau}
\label{eq-power_slow}
\ee
where
$\mathcal{P}^{(0)} = 0$ and $\mathcal{P}^{(1)} = \frac{g\kappa}{2\tau_r}\Big[(2-g)V_\infty-g\sigma^2\Big]$.

On the other hand, in the fast-cycle regime ($\tau\ll\tau_r$), the power can be expanded as a series in $\tau/\tau_r$ (Appendix B):
\be
P = P^{(0)}+P^{(1)}\frac{\tau}{\tau_r}+\mathcal{O}((\tau/\tau_r)^2),
\label{eq-power_fast}
\ee
where
\begin{subequations}
\begin{align}
P^{(0)}&=\frac{\kappa}{\tau_r}\left[V_\xi-\frac{1-(1-g)}{1+(1-g)}\sigma^2 \right],\\
P^{(1)}&=\frac{1+(1-g)}{1-(1-g)}\cdot\frac{A^2}{2\gamma}
-\frac{1+(1-g)^2}{1-(1-g)^2}P^{(0)}~.
\end{align}
\label{eq-power_fast_coeff}
\end{subequations}

A key observation from Eq. (\ref{eq-power_fast_coeff}) is that, for given $\sigma$, all the curves in Fig.~\ref{fig-maxpower_phase2}(a) share the same $P^{(0)}$, regardless of the bath activity. The effect of active noise enters only through $P^{(1)}$, which is notably independent of the active noise correlation time $\tau_c$. When $A = 0$ (no active noise), the sign of $P^{(1)}$ is always opposite to that of $P^{(0)}$ within the range $g \in (0, 2)$.  

For the standard full-shift feedback (FSF, $g=1$), Eq. (\ref{eq-power_fast_coeff}) simplifies to
\begin{subequations}
\begin{align}
P^{(0)}&=\frac{\kappa}{\tau_r}\left(V_\xi-\sigma^2\right)
=\frac{\kB T-\kappa\sigma^2}{\tau_r},\\
P^{(1)}&=
\frac{(\alpha-1)\kB T+\kappa\sigma^2}{\tau_r}.
\label{eq-power_fast_coeff_FSF}
\end{align}
\end{subequations}
where $\alpha\equiv A^2/2\kappa\kB T$. Under FSF, the sign of $P^{(1)}$ becomes positive when the active noise amplitude exceeds the threshold $A^*=\sqrt{2\kappa\kB T}$. This motivates defining the activity as a dimensionless parameter $\alpha \equiv A^2/2\kappa\kB T$.
Above the activity threshold ($\alpha>1$), the power profile exhibits a positive slope at $\tau=0$, resulting in a finite-$\tau$ maximum under FSF. The active noise correlation time determines the location and height of this maximum, as illustrated in Fig. \ref{fig-maxpower_phase2}(a).
When $\alpha<1$, a critical measurement accuracy arises, given by $\sigma^*_\alpha=\sqrt{(1-\alpha)V_\xi}$, which determines the sign of $P^{(1)}$. If $\sigma$ is greater (less) than $\sigma^*_\alpha$, $P^{(1)}$ is positive (negative) and so power grows (decays) with $\tau$.

Adjusting the feedback gain $g$ away from $1$ shifts this critical behavior. For a general feedback gain $g$, the critical measurement error $\sigma^*_\alpha(g)$, shown by the yellow boundary curves in Fig. \ref{fig-maxpower_phase2}(b), is obtained by setting $P^{(1)}=0$ in Eq.~(\ref{eq-power_fast_coeff}b). This yields
\be
\sigma_\alpha^{*2}(g)=V_\xi\left(\frac{2}{g}-1\right)
\left[1-\frac{\alpha(2-g)^2}{1+(1-g)^2 } \right]. 
\label{eq-crit_error}
\ee
due to $\kappa \tau_{\rm r}/\gamma = 1$ defined in Eq. (\ref{eq-TrajRelx}). An intriguing bifurcation occurs at $\alpha=1/2$. Below this activity level, $\sigma^*_\alpha(g)$ diverges in the limit $g\to 0$. However, above this threshold, the critical noise level suddenly collapses to zero in the same limit.

It is straightforward to see that the right-hand side of Eq.~(\ref{eq-crit_error}) remains nonnegative, regardless of $g$, provided that $\alpha\le 1/2$. However, when the activity exceeds this threshold ($\alpha>1/2$), a critical gain $g^*_\alpha$ emerges below which the expression becomes negative. Since $\sigma^2$ must be positive by definition, a negative $\sigma^*_\alpha$ indicates that the sign change of $P^{(1)}$ does not exist when $g < g^*_\alpha$. In Fig.~\ref{fig-maxpower_phase2}(b), this cutoff corresponds to the nontrivial $g$-intercept (distinct from $g=2$) of the yellow boundary curves, which appears only in the panels with $\alpha\geq 1/2$. The analytical expression for $g^*_\alpha$ is obtained by solving $\sigma_\alpha^{*2}(g)=0$, yielding
\be
g^*_{\alpha}=\frac{(2\alpha-1)-\sqrt{2\alpha-1}}{\alpha-1}, \quad\alpha\geq 1/2.
\label{eq-crit_gain}
\ee
This expression also holds for $\alpha=1$ when interpreted in the limiting sense, giving $g^*_{\alpha=1}=1$. Moreover, $g^*_\alpha$ increases monotonically from 0 at $\alpha=1/2$ and approaches 2 as $\alpha\to\infty$ implying that $g^*_\alpha<2$ for any finite activity level. 

Thus far, we have discussed the transition in the sign of $P^{(1)}$ as modulated by the measurement noise $\sigma$ for fixed feedback gain $g$. Interestingly, a similar transition occurs in the transverse direction when varying $g$ at  fixed $\sigma$. Specifically, for a given noise level $\sigma$, if the activity is less than $4/9$, the sign of $P^{(1)}$ changes only once as $g$ varies. 
However, for $\alpha > 4/9$, as detailed in Appendix~\ref{App_Shape}, the curve $\sigma^{*2}_\alpha(g)$ develops both a local maximum and a local minimum. These extrema separate further as $\alpha$ increases, until the minimum reaches zero and disappears at $\alpha=1/2$. In this intermediate range ($4/9<\alpha<1/2$), most values of $\sigma$ correspond to a single sign change in $P^{(1)}$ over $g\in(0,2)$. However, when $\sigma$ lies between the extrema, three sign reversals  occur, as illustrated in Fig.~\ref{fig-maxpower_phase2}(b). For {$\alpha>1/2$, if $\sigma$ exceeds the global maximum of $\sigma^*_\alpha(g)$, the sign of $P^{(1)}$ remains positive for all $g$. 

In summary, for $0<g<2$, while $P^{(1)}$ is always negative in a thermal bath, active noise can reverse its sign without affecting $P^{(0)}$, thereby enabling optimal power extraction at a finite cycle time $\tau$. The correlation time of the active noise, $\tau_c$, and its interplay with $\tau$ plays a crucial role in determining this optimum. In addition, even when $P^{(1)}$ is not positive, active noise can still enhance power compared to the case without active noise, as shown in Fig.~\ref{fig-maxpower_phase2}(a).

\section{Discussion and Conclusion} \label{Sec:Conclusion}
In this work, we have demonstrated that the information engines employing standard FSF encounter systematic limitations in active environments, chiefly due to the loss of system memory during the feedback process. To mitigate this issue, we introduced a proportional adjustment feedback protocol that retains a fraction of the system's memory, thereby enhancing the reliability and efficiency of work extraction.
Using a harmonic potential within an overdamped Langevin framework, we identified the covariance between the particle's position and the active noise as a key quantifier of memory effects relevant to energy extraction. While the current analysis is specific to harmonic confinement and overdamped dynamics, the proposed adjustment strategy offers a promising path forward for developing feedback schemes that outperform FSF in more complex settings, including anharmonic potentials and underdamped regimes.

Our findings underscore the importance of the feedback-relaxation interplay in the design of optimal information engines.
The proportional adjustment framework may be further extended to more general feedback strategies, including finite-time protocols\,\cite{olsen2025harnessing, garcia2025optimal, schuttler2025active}. 
However, such generalizations require a consistent and thermodynamically sound formulation of work and heat in active systems, where conventional definitions are no longer valid.
Despite these challenges, our findings provide guiding principles for designing resetting processes that exist under either thermal or active noise, or both \cite{gueneau2023active, pal2024active}.
A related theoretical framework of partial resetting \cite{olsen2024thermodynamic} may provide a complementary perspective on the efficiency of feedback-driven engines.

In this study, we assumed a Gaussian measurement error with known magnitude. In more realistic scenarios, however, the magnitude of the error may be unknown. In such cases, Bayesian inference methods, which have been successfully applied in thermal systems \cite{saha.2022}, could be employed to estimate both the magnitude of error and the optimal feedback gain. Whether these inference-based approaches can be integrated with proportional feedback in active baths remains an intriguing direction for future research.

We also examined the optimal conditions for power and found that the power output is maximized at short, finite cycle times. Hence, identifying the optimal operating time is crucial for maximizing performance. These findings offer valuable insights for developing adaptive and dynamically tunable feedback strategies in information engines, complementing recent advances in related research areas\,\cite{cocconi2024efficiency, rafeek2024active,  cocconi2025mechanical}.

\acknowledgments
This work was supported by the National Research Foundation of Korea Grants RS-2023-00263411 and RS-2024-00353098 funded by the Ministry of Science and ICT, and individual KIAS Grants No. PG064902 (J.S.L.) at the Korea Institute for Advanced Study. We gratefully acknowledge the UNIST Supercomputing Center for the support of computing resources.

\appendix
\renewcommand{\thefigure}{S\arabic{figure}}
\setcounter{figure}{0} %

\section{Derivation of (co)variance dynamics} \label{App_covariance_dynamics}
To derive the covariance dynamics, we begin by recalling the dynamics of $y_t$ for $0<t<\tau$:
\be
y_t = y_{0_+} e^{-t/\tau_r} + \frac{1}{\gamma}\int_{0_+}^t \d t'\, e^{-(t-t')/\tau_r}(\xi_{t^\prime} + \eta_{t^\prime}),
\label{eq-StocTraj_app}
\ee
where $y_{0_+}$ is the initial position immediately after feedback. Multiplying both sides by $\eta_t$ and taking the ensemble average yields the covariance
\be
C_t=\langle y_{0_+}\eta_t\rangle e^{-t/\tau_r} + \frac{1}{\gamma}\int_{0_+}^t \d t'\, e^{-(t-t')/\tau_r}\langle\eta_t\eta_{t'}\rangle.
\label{eq-cov_pre}
\ee
Our goal is to express $\langle y_{0_+}\eta_t\rangle$ in terms of $C_t$, establishing a self-consistent relation. This is possible for a broad class of active noise models, including active Ornstein–Uhlenbeck Processes (AOUP), run-and-tumble particles (RTP), and others with exponentially correlated fluctuations\,\cite{lee2022effects}.
For AOUPs, the active force evolves $\eta_t$ as
\be
\frac{1}{\tau_c}\frac{\d}{\d t} \eta_t = -\eta_t + A\sqrt{\frac{2}{\tau_c}} \zeta_t,
\label{eq-AOUP_eq}
\ee
where $\zeta_t$ is Gaussian white noise, independent of thermal noise $\xi_t$. This equation can be solved explicitly as
\be
\eta_t = \eta_0 e^{-t/\tau_c} + A \sqrt{\frac{2}{\tau_c}} e^{-t/\tau_c}\int_0^t \d t' e^{t'/\tau_c} \zeta_{t'}.
\label{eq-AOUP_dyn}
\ee
Multiplying Eq. (\ref{eq-AOUP_dyn}) by $y_{0_+}$ and taking the ensemble average gives
\be
\langle y_{0_+}\!\eta_t \rangle = C_{0_+} e^{-t/\tau_c},
\label{eq-two_point}
\ee
where $C_{0_+}=\lpa y_{0_+}\!\eta_0\rpa$. Substituting Eq. (\ref{eq-two_point}) into Eq. (\ref{eq-cov_pre}) and using the AOUP autocorrelation $\langle\eta_t\eta_{t'} \rangle=A^2\exp(-|t'-t|/\tau_c)$, we obtain the relaxation equation for the covariance $C_t$ as presented in Eq. (\ref{eq-CovRelx}) of the main text. Alternative models of active noise, such as the run-and-tumble process, also yield the same form of two-point correlation, $\langle y_{0_+} \eta_t \rangle = C_{0_+} e^{-t/\tau_c}$, ensuring consistency with the derived covariance dynamics in Eq. (\ref{eq-CovRelx}).

To obtain the variance dynamics, we first take the ensemble averages of the first and second moments of the stochastic trajectory $y_t$. From Eq. (\ref{eq-StocTraj_app}),
\begin{subequations}
\begin{align}
\langle y_t\rangle &=\langle y_{0_+}\rangle e^{-t/\tau_r}
\label{eq-First_moment}\\
\lpa y_t^2\rpa &=\left[\langle y_{0_+}^2\rangle+\frac{2}{\gamma}I_{y\eta}+\frac{1}{\gamma^2}(I_{\xi\xi}+I_{\eta\eta})\right]e^{-2t/\tau_r}.
\label{eq-Second_moment}
\end{align}
\end{subequations}
where $I_{y\eta}$ can be calculated by using Eq. (\ref{eq-two_point}):
\[
I_{y\eta}=\int_{0_+}^t \d t'\, e^{t'/\tau_r}\langle y_{0_+}\!\eta_{t'}\rangle
=C_{0_+}\frac{e^{(1/\tau_r-1/\tau_c)t}-1}{1/\tau_r-1/\tau_c}.
\] 
This result remains valid for $\tau_c=\tau_r$ under a limiting interpretation. Now using $\langle\xi_t\xi_{t'}\rangle=2\gamma \kB T \delta(t-t')$ and $\langle\eta_t\eta_{t'}\rangle=A^2\exp(-|t'-t|/\tau_c])$, we obtain
\begin{align}
I_{\xi\xi}&=
\frac{\gamma^2\kB T}{\kappa}(e^{2t/\tau_r}-1),\nn\\
I_{\eta\eta} &=
\frac{\gamma^2C_\infty}{\kappa}\left[(e^{2t/\tau_r}-1)
-2\frac{e^{(1/\tau_r-1/\tau_c)t}-1}{1-\tau_r/\tau_c}\right].\nn
\end{align}
Hence, the final expression for the variance becomes
\begin{align}
V_t &= V_{0_+}e^{-2t/\tau_r}
+\kappa^{-1}(\kB T+C_\infty)(1-e^{-2t/\tau_r})\nn \\
&\quad -2\kappa^{-1}(C_\infty-C_{0_+})
\frac{e^{-2t/\bar{\tau}}-e^{-2t/\tau_r}}{1-\tau_r/\tau_c},
\label{eq-Var_appendix}
\end{align}
which again holds in the case $\tau_r=\tau_c$ when interpreted in the limiting sense. This proves Eq. (\ref{eq-VarRelx}) in the main text.
Importantly, since Eq. (\ref{eq-two_point}) holds for a wide range of active noise models, including AOUPs, RTPs, and others with exponential autocorrelation, this derivation of the variance dynamics is broadly applicable to active baths.

\section{Asymptotic behavior of work and power in the fast and slow cycle limits}
In this section, we derive Eqs. (\ref{eq-power_slow}), (\ref{eq-power_fast}) and (\ref{eq-power_fast_coeff}), which describe the extracted power in the fast-cycle ($\tau\ll\tau_r$) and slow-cycle ($\tau\gg\tau_r$) regimes through asymptotic expansions. 
We begin by expanding the position-active noise covariance before and after feedback $C_{\tau_\pm}$, and the variance $V_{\tau_\pm}$. The structures of Eq. (\ref{eq-VarRelx}) and (\ref{eq-CovRelx}) motivate us to write as
\be
\begin{cases}
~C_{\tau_\pm}=C^{(0)}_\pm + C^{(1)}_\pm\bar\varepsilon + \mathcal{O}(\bar{\varepsilon}^2),~\bar{\varepsilon}\ll 1\\[2mm]
~V_{\tau_\pm}=V_\pm^{(0)} + V_\pm^{(1)}\varepsilon + V_\pm^{(2)}\varepsilon^2 + \mathcal{O}(\varepsilon^3),~\varepsilon \ll 1
\label{eq-CV_fast}
\end{cases}
\ee
where $\bar\varepsilon\equiv\tau/\bar{\tau}$ and $\varepsilon\equiv\tau/\tau_r$ serve as small expansion parameters in the same order.

The coefficients are determined by equating the left- and right-hand sides of Eq. (\ref{eq-CovRelx}):
\be
\begin{cases}
~C_{-}^{(0)}=C_{+}^{(0)}, \\[2mm]
~C_{-}^{(1)}=C_{+}^{(1)}-2C_{+}^{(0)}+2C_\infty.
\end{cases}
\ee
Combining these relations with
Eq. (\ref{eq-CovFeed}), we find
\be
C^{(0)}_-=0;\quad C^{(1)}_-=2g^{-1}C_\infty. 
\label{eq-C_fast_coefficient}
\ee
Similarly, by equating the left- and right-hand sides of Eq. (\ref{eq-VarRelx}), we obtain
\be
\begin{cases}
~V_{-}^{(0)}= V_{+}^{(0)},\\[2mm]
~V_{-}^{(1)}= V_{+}^{(1)}-2V_{+}^{(0)}+2V_\infty+\displaystyle\frac{2}{\kappa} \left(C_{+}^{(0)}-C_\infty\right),\\[2mm]
~V_{-}^{(2)}= V_{+}^{(2)}-2V_{+}^{(1)}+2V_{+}^{(0)}-2V_\infty+\\[1mm]
\displaystyle\frac{1}{\kappa}\left[C_{+}^{(1)} \left(1+\frac{\tau_r}{\tau_c}\right)\!-\! \left(3+\frac{\tau_r}{\tau_c}\right)\left(C_{+}^{(0)}-C_\infty\right)\right].
\end{cases}
\ee
Together with Eq. (\ref{eq-VarFeed}), this yields
\be
\begin{cases}
~V^{(0)}_{-}=V^{(0)}_{+}=\displaystyle\frac{g \sigma^2}{2-g},\\[2mm]
~V^{(1)}_{-}=\displaystyle\frac{2}{g(2-g)} \left( V_\xi - V^{(0)}_{-} \right),\\[2mm]
~V^{(2)}_{-}=\displaystyle\frac{A^2}{\kappa^2} \frac{1}{g^2} - 2 \left( V_\xi -V^{(0)}_{-} \right) \frac{1+(1-g)^2}{g^2(2-g)^2}.
\label{eq-V_fast_coeff}
\end{cases}
\ee
The work and power can then be expanded as
\begin{subequations}
\begin{align}
\langle W_\text{cyc} \rangle &= \frac{\kappa}{2}g(2-g) \left(V_{-}^{(1)}\varepsilon + V_{-}^{(2)}\varepsilon^2 \right) + \mathcal{O}(\varepsilon^3),\\
P &= \frac{\kappa}{2\tau_r}g(2-g) \left(V_{-}^{(1)} + V_{-}^{(2)}\varepsilon \right) + \mathcal{O}(\varepsilon^2).
\end{align}
\end{subequations}
These expressions reproduce Eqs. (\ref{zeropower}), (\ref{eq-power_fast}) and (\ref{eq-power_fast_coeff}).

\begin{figure}[t!] 
\centerline{
{\resizebox{\columnwidth}{!}{\includegraphics{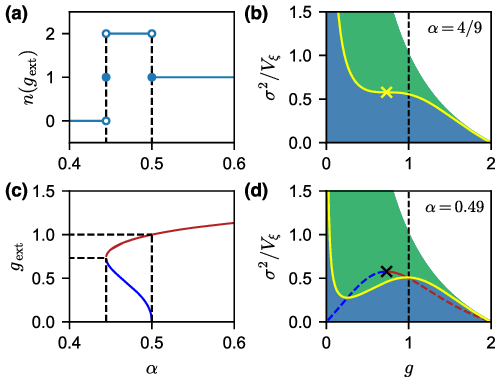}}}}
\caption{(Color online) Local extrema of $\sigma_\alpha^{*2}(g)$. (a) Number of solutions $n(g_{\rm ext})$ to $f'_\alpha(g)=0$ as a function of $\alpha$ in the regime $g \in (0,2)$. Two separate transitions occur at $\alpha=4/9$ and $\alpha=1/2$.
(b) Sign of $P^{(1)}$ at $\alpha=4/9$, analogous to Fig.~\ref{fig-maxpower_phase2}(b). The yellow curve marks $P^{(1)}=0$, and the yellow cross ($\times$) indicates its slope vanishes. 
(c) Location of the extremum $g_{\rm ext}$ as a function of $\alpha$. At $\alpha = 4/9$, a single point bifurcates into a local minimum (blue) and a local maximum (red curve). 
(d) Sign of $P^{(1)}$ at $\alpha=0.49$. The yellow curve corresponds to $P^{(1)}=0$, showing both a local minimum and a local maximum. The blue (red) curve traces the trajectory of the minimum (maximum), starting from the black cross ($\times$) at $\alpha=4/9$. For $\alpha > 1/2$, the minimum disappears while the maximum becomes the global maximum, as illustrated in Fig.~\ref{fig-maxpower_phase2}(b), particularly in the panels for $\alpha=1/2$, $0.52$, and $1$.}
\label{fig-power_appendix}
\end{figure}

For the slow-cycle regime, we expand
\be
\begin{cases}
~C_{\tau_\pm}= 
\mathcal{C}^{(0)}_\pm + \mathcal{C}^{(1)}_\pm \bar{\varepsilon} + \mathcal{O}(\bar{\varepsilon}^2),~\bar{\varepsilon} \gg 1\\[2mm]
~V_{\tau_\pm}= \mathcal{V}_\pm^{(0)} + \mathcal{V}_\pm^{(1)}\varepsilon + \bar{\mathcal{V}}_\pm^{(1)}\bar{\varepsilon} + \mathcal{O}(\varepsilon^i\bar{\varepsilon}^j),~\varepsilon \gg 1
\label{eq-CV_slow}
\end{cases}
\ee
where $\varepsilon = e^{-2\tau/\tau_r}$ and $\bar{\varepsilon}=e^{-2\tau/\bar{\tau}}$ serve as small expansion parameters of comparable order. From Eq. (\ref{eq-CovRelx}), we have
\be
\mathcal{C}^{(0)}_- = C_\infty; \quad 
\mathcal{C}^{(1)}_- = -g C_\infty. 
\label{eq-C_slow_coefficient}
\ee
Likewise, from Eq. (\ref{eq-VarRelx}) we obtain
\be
\begin{cases}
~\mathcal{V}^{(0)}_{-}= V_\infty\\[2mm]
~\mathcal{V}^{(1)}_{-}= g^2 \sigma^2 - g(2-g) V_\infty + \displaystyle\frac{2gC_\infty}{\kappa(1-\tau_r/\tau_c)}\\[2mm]
~\bar{\mathcal{V}}^{(1)}_{-}=\displaystyle\frac{-2gC_\infty}{\kappa(1-\tau_r/\tau_c)}.
\label{eq-V_slow_coeff}
\end{cases}
\ee
The work per cycle then follows as
\begin{align}
\langle W_\text{cyc} \rangle 
&=W_\infty -g(2-g) W_\infty \varepsilon +g^2(2-g) C_\infty \frac{\varepsilon-\bar{\varepsilon}}{1-\tau_r/\tau_c}\nn\\
&\quad+\mathcal{O}(\varepsilon^i \bar{\varepsilon}^j_{i+j \geq 2}),
\label{Wcyc}
\end{align}
where $W_\infty \equiv \frac{\kappa}{2} (\mathcal{V}_{-}^{(0)}-\mathcal{V}_{+}^{(0)})=\lim_{\tau \to \infty} \langle W_\text{cyc} \rangle$. This result reproduces Eqs. (\ref{eq-work_slow_limit}) and (\ref{eq-power_slow}). Note that Eq. (\ref{Wcyc}) also holds in the case $\tau_r=\bar{\tau}$, when interpreted in the limiting sense.

\section{Local extrema of the $\sigma_\alpha^{2}(g)$ curve} \label{App_Shape}
From Eq.~\eqref{eq-crit_error}, we define 
\be
f_\alpha(g)\equiv \sigma_\alpha^{*2}(g)/V_\xi.
\ee 
The function $f_\alpha(g)$ and its derivative with respect to $g$ are given by
\be
\begin{cases}
~f_\alpha(g)=\displaystyle\frac{N_\alpha^{(0)}(g)}{g\left(1+(1-g)^2 \right)},\\[3mm]
~f^\prime_\alpha(g)=\displaystyle\frac{N_\alpha^{(1)}(g)}{g^2 \left( 1+(1-g)^2 \right)^2},
\end{cases}
\ee
where $N_\alpha^{(0)}(g)$ and $N_\alpha^{(1)}(g)$ are 
\begin{align}
N_\alpha^{(0)}(g) &= (\alpha-1)g^3 + (4-6\alpha)g^2 + (12\alpha-6)g \nn \\ &\quad+ (4-8\alpha), \nn \\
N_\alpha^{(1)}(g) &= (4\alpha-2)g^4 + (8-20\alpha)g^3 + (36\alpha-16)g^2 \nn \\ &\quad+ (16-32\alpha)g + (16\alpha-8). \nn
\end{align}
For $\alpha=1/2$, the limit $\lim_{g \to 0} f_\alpha(g)$ converges to the finite value $0$. Fig.~\ref{fig-power_appendix}(a) shows the number of solutions to $N_\alpha^{(1)}(g)=0$ for $g\in (0,2)$, i.e., the number of local extrema of $\sigma_\alpha^{2}(g)$. For $\alpha < 4/9$, no solutions exist and thus $n(g_{\rm ext})= 0$, where $g_{\rm ext}$ denotes the solution of $N_\alpha^{(1)}(g) =0$, i.e., the location of an extremum. At $\alpha = 4/9$, a single solution appears, as shown in Fig.~\ref{fig-power_appendix}(b). For $4/9<\alpha<1/2$, both a local minimum (blue curve in Figs.~\ref{fig-power_appendix}(c) and (d)) and a local maximum (red curve in Figs.~\ref{fig-power_appendix}(c) and (d)) emerge, giving $n(g_{\rm ext})=2$ up to $\alpha = 1/2$. For $\alpha>1/2$, the local minimum disappears, leaving only the maximum, so $n(g_{\rm ext})=1$. 

\newpage
%

\end{document}